\documentclass[10pt]{iopart}
\usepackage[super,compress]{cite}
\usepackage{graphicx}
\usepackage{subfigure}
\expandafter\let\csname equation*\endcsname\relax
\expandafter\let\csname endequation*\endcsname\relax
\usepackage{amsmath}
\usepackage{mathtools}
\usepackage{iopams}
\usepackage{amssymb}
\usepackage{breakurl}
\usepackage{url,hyperref}
\usepackage{float}
\usepackage{multirow}
\def\amu{$\Delta{a^{\rm FB}_\mu}$}
\newcommand{\code}[1]{\texttt{#1}}

\newcommand{\beqn}{\begin{eqnarray}}
\newcommand{\eeqn}{\end{eqnarray}}
\newcommand{\be}{\begin{equation}}
\newcommand{\ee}{\end{equation}}

\def \n34{\tilde{\chi}^{0}_{3,4}}

\begin{document}

\begin{flushright}
MS-TP-21-39
\end{flushright}

\title[A. Aboubrahim, M. Klasen, P, Nath and R. M. Syed]{Tests of gluino-driven radiative breaking of the electroweak symmetry at the LHC}

\author{Amin Aboubrahim$^1$, Michael Klasen$^1$, Pran Nath$^2$ and Raza M. Syed$^3$}
\address{$^1$ Institut f\"ur Theoretische Physik, Westf\"alische Wilhelms-Universit\"at M\"unster, \\ Wilhelm-Klemm-Stra{\ss}e 9, 48149 M\"unster, Germany} 
\vspace{1mm}
\address{$^2$ Department of Physics, Northeastern University, Boston, MA 02115-5000, USA}
\vspace{1mm}
\address{$^3$ Department of Physics, American University of Sharjah, \\ P.O. Box 26666, Sharjah, UAE}

\ead{aabouibr@uni-muenster.de, michael.klasen@uni-muenster.de, p.nath@northeastern.edu, rsyed@aus.edu}
\vspace{10pt}

\begin{abstract}

The recent muon $g-2$ result from Fermilab combined with the Brookhaven result, strongly points to new physics beyond the Standard Model which can be well described by the electroweak sector of supersymmetry if the masses of the sleptons and some of the 
electroweak gauginos are in the few hundred GeV range. 
However, the Higgs boson mass measurement at 125 GeV indicates
 a mass scale for squarks which lies in the few TeV region indicating
 a split mass spectrum between squarks and sleptons. This apparent
 puzzle is resolved in a natural way in gluino-driven radiative breaking
 of the electroweak symmetry where radiative breaking is driven 
 by a large gluino mass and the gluino color interactions lead to 
 a large splitting between the squarks and the sleptons. We show
 that an analysis without prejudice using  an artificial neural network also
 leads to the gluino-driven radiative breaking.
   We use  a set of benchmarks and a deep neural network analysis
    to test the
   model for the discovery of  light sleptons and sneutrinos 
   at HL-LHC and HE-LHC.

\end{abstract}

\vspace{2pc}

\submitto{\PS}
\maketitle
%\ioptwocol
%\tableofcontents 

\section{Introduction}	

In the Standard Model (SM), the muon anomalous magnetic moment is computed to a very high precision and is predicted to have the value~\cite{Aoyama:2020ynm,Davier:2019can,Davier:2017zfy,Davier:2010nc,Crivellin:2020zul,Keshavarzi:2020bfy,Colangelo:2020lcg}   
 \begin{equation}
 a^{\rm SM}_\mu = 116 591 810 (43)  \times10^{-11}.
  \label{SM}
\end{equation}
The recent measurement of the muon $g-2$ from Fermilab~\cite{Abi:2021gix} combined with the older Brookhaven~\cite{Bennett:2006fi,ParticleDataGroup:2018ovx} measurement gives
 \begin{equation}
 a^{\rm exp}_\mu = 116 592 061 (41)  \times10^{-11}.
  \label{Brook}
\end{equation} 
The difference between the SM and experimental results is 
  \begin{equation}
\Delta a^{\rm FB}_\mu = a_{\mu}^{\rm exp} - a_{\mu}^{\rm SM} = 251 (59)  \times10^{-11},
  \label{diff}
\end{equation} 
which corresponds to a $4.1\sigma$ deviation from the SM. This result brings about a confirmation of the previous Brookhaven observation which stood at $3.7 \sigma$ significance (see, however, Ref.~\cite{Borsanyi:2020mff}).
The new result constitutes an update to the constraints on supergravity (SUGRA) unified models and its implication on supersymmetry (SUSY) was analyzed in Ref.~\cite{Aboubrahim:2021rwz}. 
It has been known for some time now~\cite{Kosower:1983yw,Yuan:1984ww}  that a supersymmetric electroweak correction to the muon $g-2$ can compete with the ones from the SM electroweak sector. Furthermore, the muon anomaly is sensitive to the soft parameters in SUGRA models~\cite{Lopez:1993vi,Chattopadhyay:1995ae,Moroi:1995yh} which affect the sparticle masses which enter in the loop corrections and contribute to the muon anomaly. Specifically,  the smuon and muon-sneutrino as well as the chargino and the neutralino must be light with masses lying in the sub-TeV region so that their loop corrections to the muon $g-2$ anomaly can be sizable. Thus, within SUGRA high scale models, the muon anomaly seems to favor smaller universal scalar masses. On the other hand, we know that the discovery of the Higgs boson near $\sim 125$ GeV~\cite{Aad:2012tfa,Chatrchyan:2012ufa} indicates that the size of weak scale SUSY lies in the several TeV region and a large scalar mass would be required to generate a large loop correction to raise the tree level Higgs mass from just below the $Z$ boson mass to the experimentally observed value~\cite{Akula:2011aa,higgs7tev1}. This seems to go against the requirement for a sizable contribution to the muon $g-2$ from SUSY. We discuss here a possible solution to this discrepancy where a gluino-driven radiative breaking of the electroweak symmetry in SUGRA can render a spectrum with the correct Higgs mass as well as light weakinos and light sleptons which can explain the $4.1\sigma$ deviation seen by Fermilab in the muon $g-2$. In this case, squarks, which carry color, acquire large masses the size of several TeV in renormalization group (RG) running while the color neutral sleptons remain light with masses around  the electroweak scale. This creates a split spectrum with a heavy colored sector and a light electroweak 
sector~\cite{Akula:2013ioa,Aboubrahim:2019vjl,Aboubrahim:2020dqw,Aboubrahim:2021phn}.  \\

The outline of the rest of the paper is as follows: In section~\ref{sec:grewsb}, we 
discuss the gluino-driven radiative breaking with specific reference to grand unified models. In section~\ref{sec:scan}, a  broad scan of the SUGRA parameter space using an artificial neural network is discussed.  Also discussed is the sparticle spectrum arising from 
the scan which exhibit a large mass gap between the squark masses and the slepton masses. 
In section~\ref{sec:split}, cosmologically consistent SUGRA benchmarks within 
$\tilde g$REWSB satisfying the constraints on relic density, Higgs boson mass,  $g-2$ and LHC sparticle production and with split squark-slepton spectra 
 are discussed. In section~\ref{sec:dnn}, 
 an analysis for the detection of sparticles using deep neural network
 is given and section~\ref{sec:conc} contains the  conclusions.

\section{Gluino-driven radiative breaking in grand unified models}\label{sec:grewsb}
 
In this section we discuss how a large mass
 hierarchy between masses of the squarks
 versus those for the sleptons  can be generated in a natural fashion in SUGRA models via color interactions.  
 Let us begin with some general formulas for  squark  masses at one loop for the first two generations where we allow for
non-universalities in the soft parameter for
both the scalar masses and for the gaugino
masses. While inclusion of higher loop corrections in RG evolution
is important for precision physics, the main conclusions regarding the
generation of mass hierarchy between squarks and sleptons can be
drawn just from the consideration of one loop RG which we now 
discuss. Thus for the case of the first two generations of squarks 
where one can neglect the Yukawa coupling in the evolution, 
the squark masses for the first generations ($i=1,2$; where $i$ is a generation index) are given by 
\begin{eqnarray}
m^2_{\tilde u_{iL}} =& m^2_{\tilde q_{i0}}  +m^2_{u_i}
+ \tilde \alpha_G \left( \frac{8}{3} f_3 m_3^2+ \frac{3}{2} f_2m_2^2 + \frac{1}{30}  f_1 m_1^2\right) \nonumber\\
&+ \left(\frac{1}{2} - \frac{2}{3} \sin^2\theta_W\right) M_Z^2
\cos(2\beta), \nonumber \\
m^2_{\tilde d_{iL}} =& m^2_{\tilde q_{i0}} +m^2_{d_i}
+ \tilde \alpha_G \left( \frac{8}{3} f_3 m_3^2+ \frac{3}{2} f_2m_2^2 + \frac{1}{30}  f_1 m_1^2 \right) \nonumber\\
&+ \left(-\frac{1}{2} + \frac{2}{3} \sin^2\theta_W\right) M_Z^2
\cos(2\beta), \nonumber\\
m^2_{\tilde u_{iR}} =& m^2_{\tilde u_{i0}} +m^2_{u_i}
+ \tilde \alpha_G \left( \frac{8}{3} f_3 m_3^2+  \frac{8}{15}  f_1 m_1^2\right) \nonumber\\
&+ \frac{2}{3} \sin^2\theta_W M_Z^2
\cos(2\beta), \nonumber\\
m^2_{\tilde d_{iR}} =& m^2_{\tilde d_{i0}} +m^2_{d_i}
+ \tilde \alpha_G \left( \frac{8}{3} f_3 m_3^2+  \frac{2}{15}  f_1 m_1^2\right) \nonumber\\
&- \frac{2}{3} \sin^2\theta_W M_Z^2
\cos(2\beta).
\label{sqa-1}
\end{eqnarray}
In the above $m_{\tilde q_{i0}}$ etc.  are the soft scalar masses, $m_1,m_2,m_3$ 
are the soft gaugino masses for $U(1)$, $SU(2)$ and $SU(3)$ gauginos and  
$\tan\beta=\langle H_2\rangle/\langle H_1\rangle$,
where $\langle H_2\rangle$ gives mass to the up quarks and $\langle H_1\rangle$ gives mass to the down quarks and the leptons all taken at the grand unification scale of $M_G=2\times 10^{16}$ GeV.
Further, we have $f_k(t) = t(2-\beta_k t)/(1+\beta_k t)^2$, $\beta_k= (33/5,1, -3)$.
 With the same approximation neglecting the Yukawa couplings,   
 slepton masses for all three generations  are given by 
\begin{eqnarray}
m^2_{\tilde e_{iL}} =& m^2_{\tilde \ell_{i0}} +m^2_{e_i}
+ \tilde \alpha_G \left(\frac{3}{2} f_2m_2^2 +
 \frac{3}{10}  f_1 m_1^2\right) \nonumber\\
&+ \left(-\frac{1}{2} +\sin^2\theta_W\right) M_Z^2
\cos(2\beta), \nonumber\\
m^2_{\tilde \nu_{iL}}=&m^2_{\tilde \nu_{i0}}
+ \tilde \alpha_G \left( \frac{3}{2} f_2m_2^2+ \frac{3}{10}  f_1 m_1^2\right) 
+ \frac{1}{2} M_Z^2
\cos(2\beta), \nonumber\\
m^2_{\tilde e_{iR}} =& m^2_{\tilde e_{i0}} +m^2_{e_i}
+ \tilde \alpha_G \frac{6}{5}  f_1 m_1^2
-\sin^2\theta_W M_Z^2\cos(2\beta).
\label{lep-1}
\end{eqnarray}
Here $i=1\cdots 3$ runs over the three lepton species and $m_{\tilde \ell_{i0}}$ etc.
are in general non-universal mass parameters.
To introduce a very significant splitting between the squark and slepton 
masses one could choose the soft parameters for the squarks to be much 
larger than those for the sleptons at the GUT scale.
However, in grand unified models this is 
not possible since the quarks and the leptons enter in common group representations.
Thus, for example, in $\mathsf{SU(5)}$ the quarks and the leptons belong to the $\mathsf{\bar 5}$ and $\mathsf{10}$-plet
representations where $\mathsf{\bar 5}$ has the particle content of $(d^c_L, \nu_L, e_L)$ and 
$\mathsf{10}$-plet has the particle content $(u_L, d_L, u^c_L, e^c_L)$. In $\mathsf{SO(10)}$ a single quark-lepton
generation belongs to the $\mathsf{16}$-plet representation of $\mathsf{SO(10)}$. Thus in grand unified models,
assigning soft masses to squarks different from those for the sleptons which lie in the same 
multiplet is difficult. However, this is not the case for the gaugino masses. Here the gaugino masses, arising from $F$-type breaking, have many possibilities arising from irreducible representations in the
symmetric product of two adjoint representations of the grand unified group, i.e., one has the gaugino mass term given  by
\be
- \frac{\langle F_{ab}\rangle }{M_{\rm Pl}} \frac{1}{2} \lambda_a \lambda_b + {\rm h.c.}, 
\ee
where $\langle F_{ab}\rangle$ is a VEV of dimension 2 and $M_{\rm Pl}$ is the
Planck mass.  Thus for $\mathsf{SU(5)}$ the 
symmetric product of the adjoint representation 
$(24\times 24)_{\rm sym}$ contains 
 $\mathsf{1,24, 75, 200}$. Each of these give specific set of ratios for the $U(1), SU(2), SU(3)$ gaugino masses~\cite{Martin:2009ad,Feldman:2009zc}.
  Thus for the singlet representation one has $m_1:m_2:m_3=(1:1:1)$, for the $\mathsf{24}$-plet one has
the ratio $(-\frac{1}{2}:-\frac{1}{2}:1)$,  the $\mathsf{75}$-plet gives $(-5:3:1)$ while the $\mathsf{200}$-plet gives $(10:2:1)$.
For $\mathsf{SO(10)}$ the symmetric product of the two adjoint representations
$(45\times 45)_{\rm sym}$ contains the representations $\mathsf{210}$
and $\mathsf{770}$ where $\mathsf{210}$ 
 gives $m_1:m_2:m_3$ ratios $(-\frac{3}{5}:1:0)$,
$(-\frac{4}{5}:0:1)$ and $(1:0:0)$ while $\mathsf{770}$ gives the ratios
$(\frac{19}{10}: \frac{5}{2}: 1)$ and $(\frac{32}{5}:0:0)$.
Here we note that  a linear combination of the above allows for non-universalities for the three gaugino masses. Further,  we note that
different $m_1:m_2:m_3$ but with the same value of 
$r\equiv (m_2-m_1/(m_3-m_1)$ are isomorphic under re-definitions
and scalings in the gaugino sector.
With the mechanism in place for generating 
non-universalities~\cite{Martin:2009ad,Feldman:2009zc,nonuni2,Belyaev:2018vkl},
we assume now that $m_3\gg m_2, m_1,m_0$, where $m_0$ is the universal 
scalar mass.
 In this case, 
the largest contribution to the first two generations of squarks is dominated by the gluino mass $m_3$, and for small $m_0$ all the squark masses are typically 
the same size, i.e., 
\be
 m^2_{\tilde u_{iL}} \sim m_{\tilde d_{iL}}
\sim m^2_{\tilde u_{iR}} \sim m^2_{\tilde d_{iR}}\sim \frac{8}{3} f_3 m_3^2.
\ee 

\begin{figure}[!htp]
\begin{center}
\includegraphics[width=0.6\textwidth]{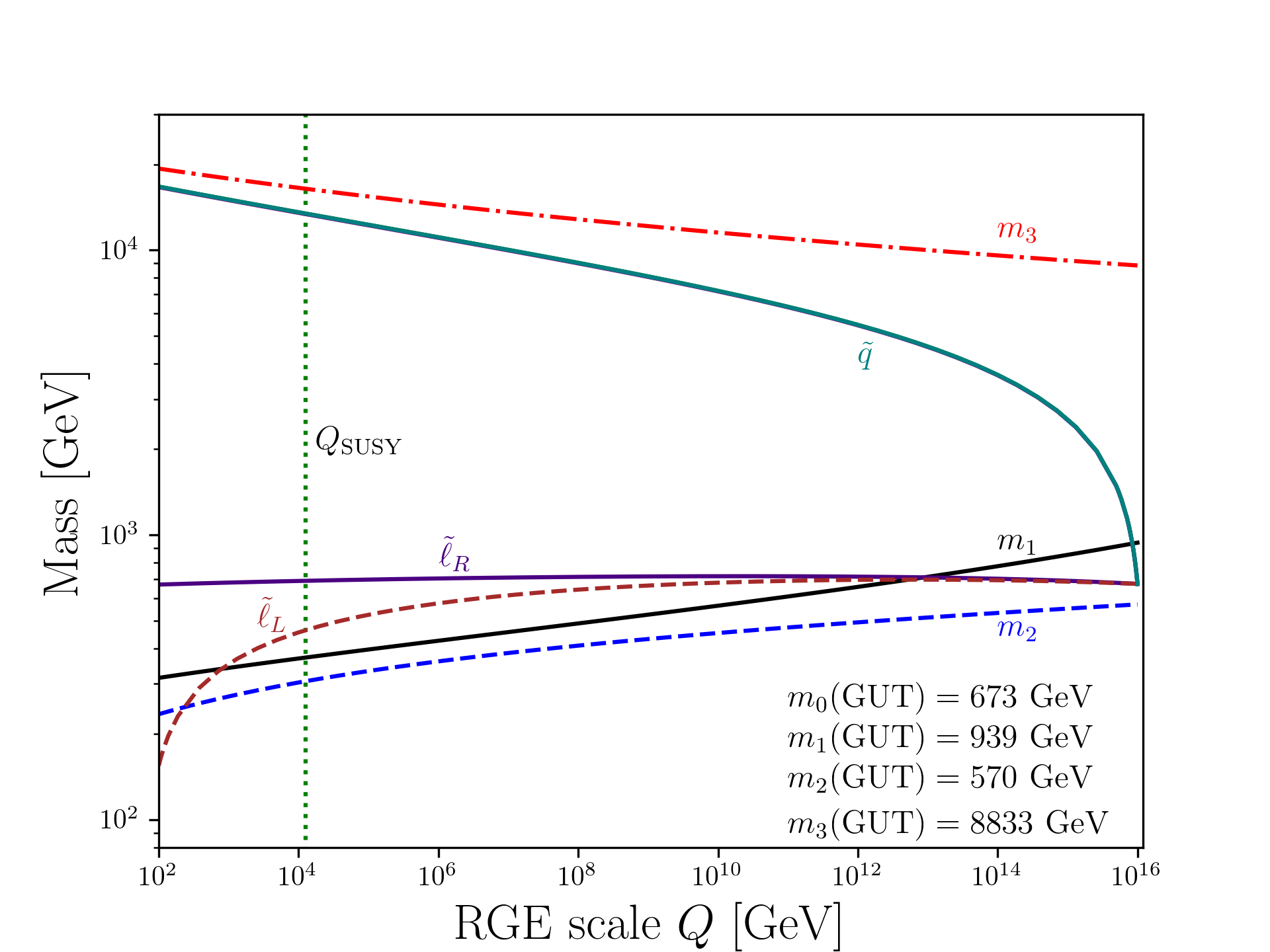}  
  \caption{\label{splitting}
  Exhibition of a natural splitting between the slepton and squark masses in the gluino-driven radiative breaking
    of the electroweak symmetry. The running of the gaugino masses is also shown. The SUSY breaking scale, $Q_{\rm SUSY}$, is defined as the geometric average of the two stop masses (after~\cite{Akula:2013ioa}).    }  
\end{center}
\label{fig:rge}
\end{figure}

For the third generation squarks, the Yukawa couplings do play
a role in RG evolution and here we have the coupled equations
\begin{eqnarray}
\frac{dm_{\tilde Q}^2}{dt}&=&-Y_t\Sigma -Y_t A^2_t
+\left(\frac{16}{3}\tilde\alpha_3m_3^2
+ 3\tilde\alpha_2m_2^2+
\frac{1}{9} 
\tilde\alpha_1m_1^2\right)~\label{dmq2},~\nonumber\\
\frac{dm_{\tilde U}^2}{dt}&=&-2Y_t\Sigma -2 Y_t A_t^2
+\left(\frac{16}{3}\tilde\alpha_3m_3^2
+ \frac{16}{9} \tilde\alpha_1m_1^2\right)~\label{dmu2},\nonumber\\
\frac{dm_{H_2}^2}{dt}&=&-3Y_t\Sigma  - 3Y_t A_t^2
+\left(3\tilde\alpha_2m_2^2+
\tilde\alpha_1m_1^2\right),\label{dmh2}
\end{eqnarray}
where $\Sigma= (m_{\tilde q}^2 + m_{\tilde u}^2+ m_{\tilde H_2}^2)$ and $Y_t=h_t^2/(4\pi)^2$, where $h_t$ is the Yukawa coupling for the top quark and $A_t$ is the trilinear coupling for the top squarks. 
Here we assume a universal scalar mass $m_0$. 
Near the grand unification scale the ratio of $Y_t \Sigma$ and of
$Y_t A_t^2$ versus $\tilde \alpha_3 m_3^2$  is controlled by
$(m_0/m_3)^2$ which is typically in the range $(10^{-2}-10^{-3})$
if $m_0$ lies in the few hundred GeV region, as needed for the sleptons to be
light, and $m_3$ is of size a few TeV 
and thus gluino dominates radiative breaking. In this case  the    
    third generation squark masses will
   again have their masses of size similar to those of the first two generation squarks. 
   
   In summary, since in grand unified models  
 the quarks and leptons (and squarks and
 sleptons) lie in common multiplets the squarks and sleptons have a common mass 
 at the GUT scale. 
  This makes it difficult
 to create  a large mass hierarchy between
 squarks and sleptons. On the other hand 
 color interactions  can generate a large
 splitting since they discriminate between
 squarks and sleptons. This is exactly what
 gluino-driven radiative electroweak symmetry breaking ($\tilde g$REWSB)  does.
 In the analysis below we carry out a broad
 probe of the SUGRA parameter space to 
 show that  $\tilde g$REWSB is indeed the
 preferred mechanism for understanding both
 the Higgs boson mass and the Fermilab muon $g-2$
 result.

\section{A scan of the SUGRA parameter space  using an artificial neural network}\label{sec:scan}

As mentioned above, the Higgs boson mass of $\sim 125$ GeV points towards a weak scale SUSY in the several TeV range. In SUGRA models, this can be in general achieved with a large universal scalar mass. However, the muon $g-2$ requires a light smuon mass which cannot be achieved with a large scalar mass. In order to investigate the SUGRA parameter space that can reconcile the two apparent conflicting constraints, we employ an artificial neural network (ANN). Without any a priori assumption on the sizes of the soft parameters, we run an ANN to scan the SUGRA parameter space while applying the constraints on the Higgs boson mass, the dark matter relic density and the recent Fermilab muon $g-2$ anomaly.   
Neural networks are efficient in analyses when one deals with a large parameter space (see, e.g., Refs.~\cite{Hollingsworth:2021sii,Balazs:2021uhg}) and this is the  case for models we investigate. Specifically, as discussed above 
 we investigate the SUGRA model with non-universalities.
The result of the scan shows that the preferred region within the 
prescribed SUGRA parameter space is one where the  universal scalar mass at the GUT scale is relatively small lying in the few hundred GeV region but the gluino mass is large and lying in the few TeV region. Owing to the large gluino mass, the RG evolution will drive the squark masses to the several TeV range while the sleptons remain light.  This mechanism then leads to
    a natural splitting between the slepton and squark masses as shown in the left panel of Fig.~\ref{splitting} as noted in the
    early work of~\cite{Akula:2013ioa}.   
     The large mass of the squarks allows one
 to satisfy the Higgs boson mass constraint, while the relative smallness  of the slepton masses allows
 us to produce a significant correction to the muon $g-2$ anomaly indicated by the new measurement.

\begin{figure}[!htp]
\begin{center}
 \includegraphics[width=0.5\textwidth]{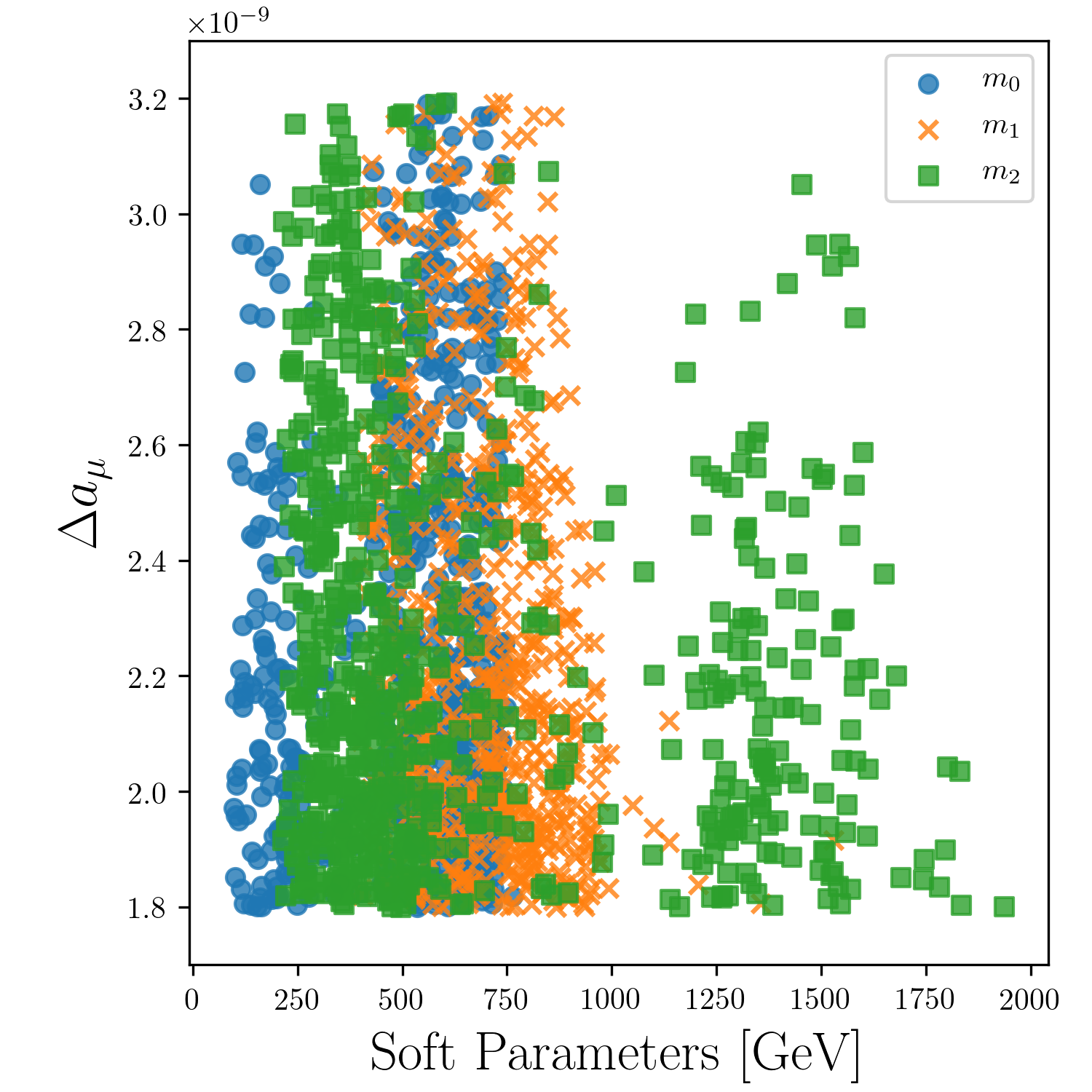}
    \caption{\label{m0m1m2}
A scatter plot of $\Delta a_{\mu}$ vs
 $m_0, m_1, m_2$   using an ANN analysis consistent with $\Delta a_\mu^{\rm FB}$ within $1\sigma$
in model with gluino driven radiative breaking
of the electroweak symmetry. 
The model points satisfy  all constraints mentioned in the text~\cite{Aboubrahim:2021ily}.
   }  
\end{center}
\end{figure}

The scan produces a large set of points which are passed to \code{Lilith}~\cite{Bernon:2015hsa,Kraml:2019sis}, \code{HiggsSignals}~\cite{Bechtle:2013xfa} and \code{HiggsBounds}~\cite{Bechtle:2020pkv} to check the Higgs sector constraints as well as \code{SModelS}~\cite{Khosa:2020zar,Kraml:2013mwa,Kraml:2014sna} to apply the sparticle mass limits from the LHC. Further, we use the module embedded in \code{micrOMEGAs-5.2.7}~\cite{Barducci:2016pcb} to check the constraints from DM direct detection experiments as well as to determine the DM relic density.
After applying all the above constraints, we display the surviving SUGRA model points as a scatter plot in the right panel of Fig.~\ref{m0m1m2} where we exhibit $\Delta a_\mu$ vs
  $m_0, m_1, m_2$.
  %%%%%%%%%%%%%%
  The analysis shows that there is a wide region of the SUGRA parameter space which is consistent within the $1\sigma$ range of $\Delta a^{\rm FB}_\mu$ rather than just a few model points. 
  
\begin{figure}[!htp]
  \begin{center}
    \includegraphics[width=0.49\textwidth]{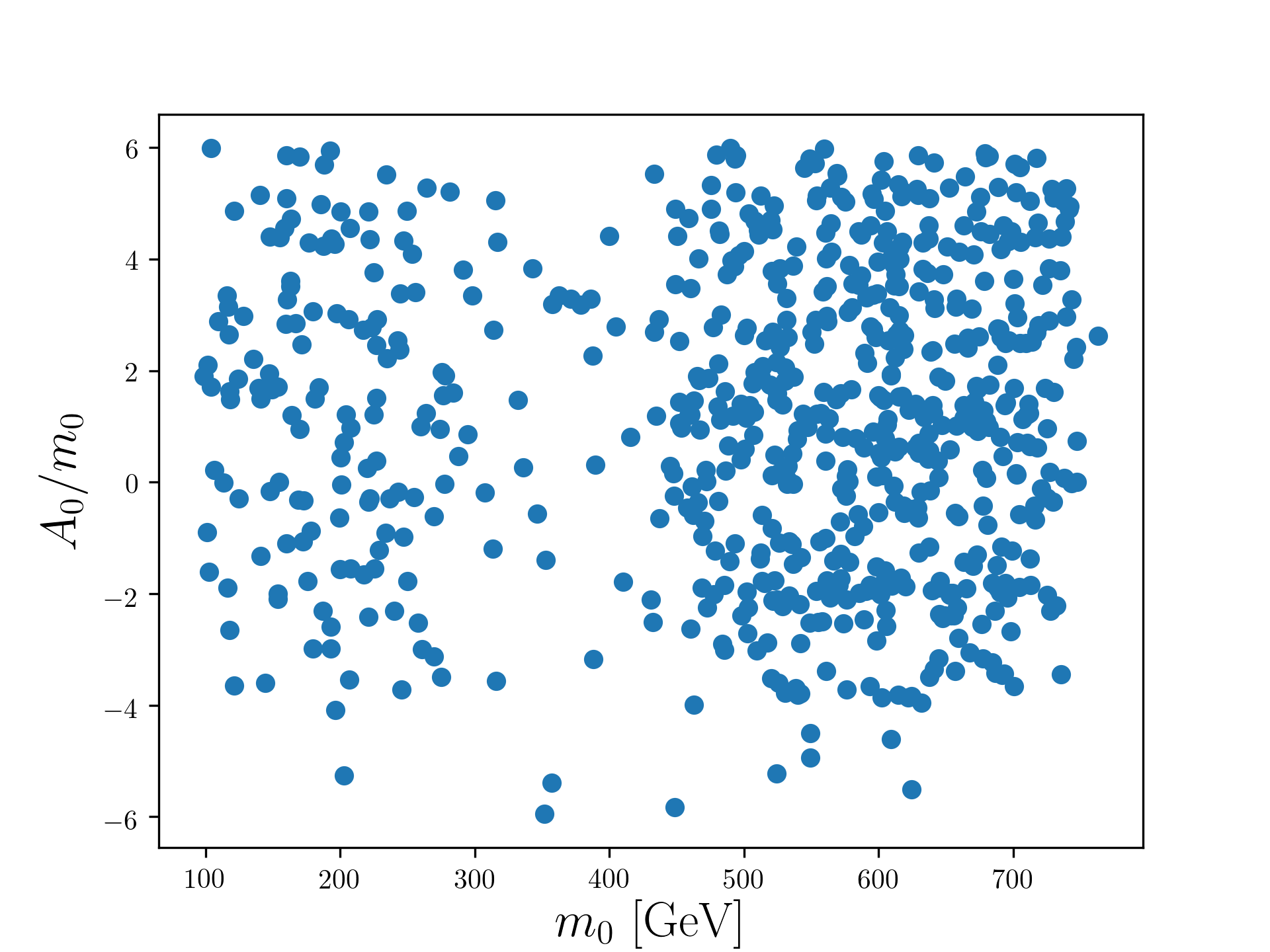}
  \includegraphics[width=0.49\textwidth]{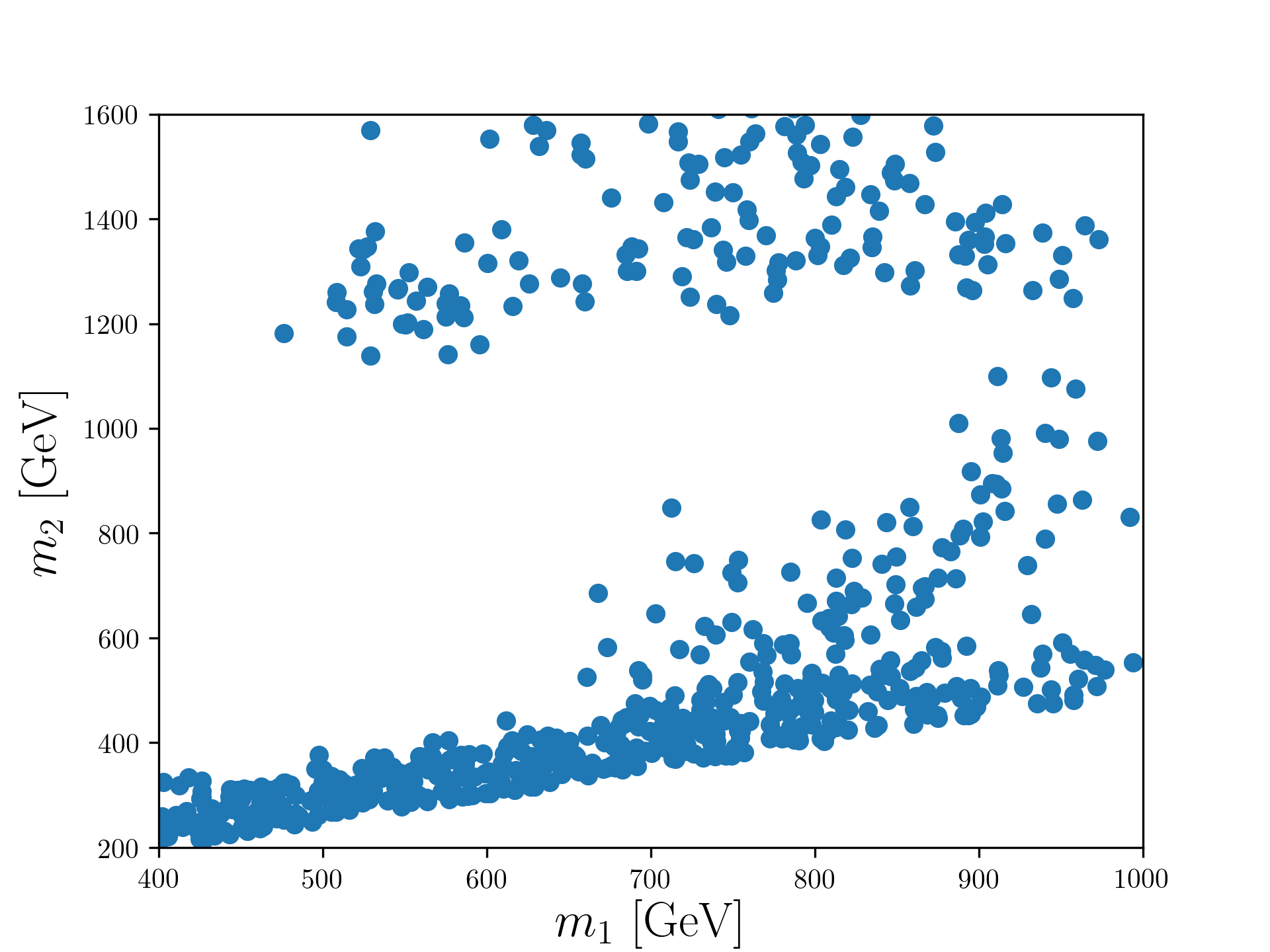}  
    \caption{\label{a0tb}
  An artificial neural network
     analysis of SUGRA  parameter space which generates the desired correction to the muon anomaly indicated
     by the Fermilab experiment.  
  Left panel: Scatter plot of $A_0/m_0$ vs. $m_0$ which shows $A_0/m_0$ in the range $(-6, 6)$ and $m_0$ in the range $(100,700)$. Specifically, a significant number of model points lie in 
  the low mass range $(100,300)$ GeV. Right panel:  A scatter plot 
  in the $m_2$ vs $m_1$ which shows a significant population of model
  points in the few hundred GeV region. }
  \label{fig3}
  \end{center}
\end{figure}

  In Fig.~\ref{a0tb} the left panel gives a scatter plot of $A_0/m_0$ versus $m_0$ 
  and here one  finds that $A_0/m_0$ lies  in the range $(-6,6)$ while
  $m_0$ lies in the range (100$-$700) GeV with many model points lying in the low mass range of  (100$-$300) GeV. The right panel gives a scatter plot 
  in the $m_2$ versus $m_1$ plane and shows a significant population of model point in the few hundred GeV gaugino mass region.  LHC constraints on sparticle masses disfavor the small $m_0$ region (left panel of Fig.~\ref{a0tb})  while a larger number of points at higher values of $m_0$ remain viable. The opposite is true for $m_2$ where larger $m_2$  are disfavored (right panel of Fig.~\ref{a0tb}). In particular, for $m_2>m_1$, the right handed slepton becomes lighter than the left handed one which means more constrained by LHC analyses.
  
\begin{figure}[!htp]
\hspace{2.4cm} {Light Spectrum} \hspace{3.4cm} {Heavy Spectrum}
  \begin{center}
  \includegraphics[width=0.49\textwidth]{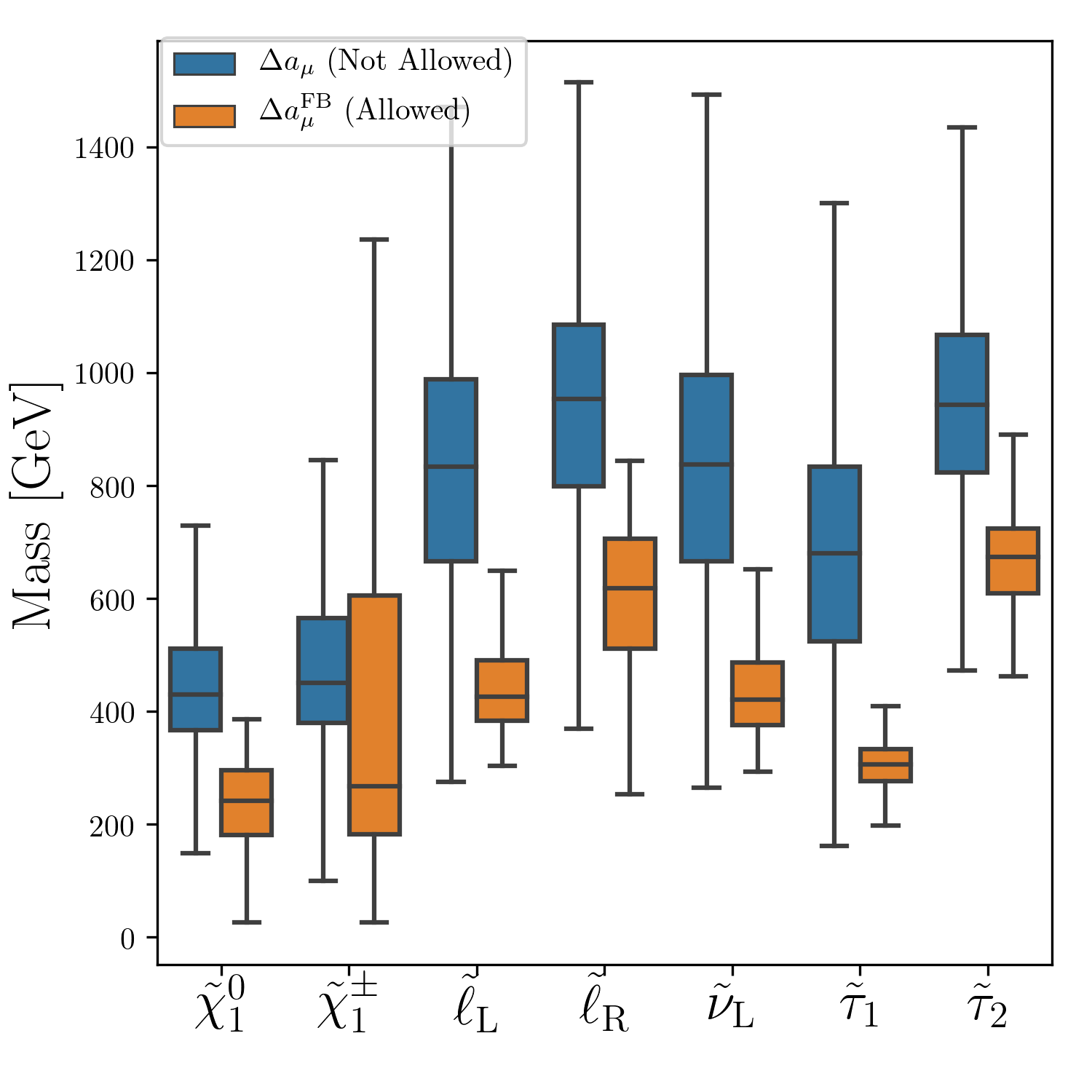}
  \includegraphics[width=0.49\textwidth]{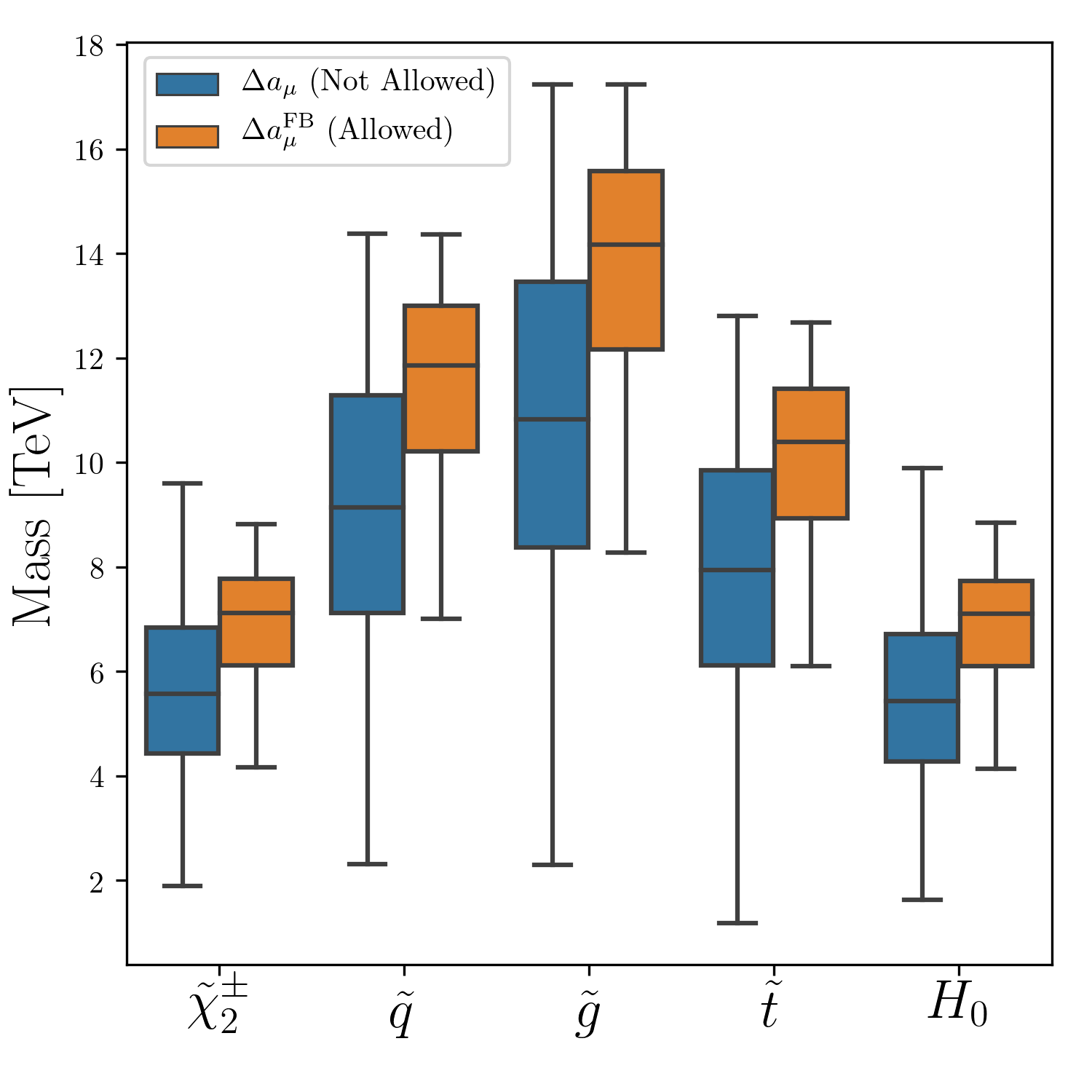}
    \caption{\label{splitmass}
   A display of the split sparticle spectrum consisting of light weakinos, sleptons, sneutrino and staus (left panel) and the heavy chargino, squarks, gluino, lightest stop and the CP even Higgs (right panel) that emerge in the
    gluino-driven radiative breaking of the electroweak symmetry in $\tilde g$SUGRA grand unified models. Each subplot shows a probability density distribution in the particles' masses for two cases: the region consistent with $\Delta a_{\mu}^{\rm FB}$ (orange) and  the region outside of the muon $g-2$ error bars (blue). } 
   \end{center}
\end{figure}

A display of the sparticle mass hierarchy 
  generated by the RG running is given in Fig.~\ref{splitmass}. Here the left panel shows the
  light sparticle spectrum while the right panel shows the heavy sparticle masses. Corrections to the muon
  $g-2$ are governed by the light spectrum of the left panel and this mass spectrum lies in the range of few hundred GeV while the heavy mass spectrum lies in the mass range of few TeV.
  Because of this split mass spectrum, the sparticles most accessible for SUSY searches at the LHC
  are the ones in the left panel.  For the low-lying spectrum consistent with $\Delta a_{\mu}^{\rm FB}$, the mean value (shown by the solid line) of the chargino mass is $\sim 445$ GeV,  the lightest neutralino mass $\sim 235$ GeV, the left handed slepton mass $\sim 480$ GeV, the right handed slepton mass $\sim 590$ GeV, the sneutrino mass $\sim 470$ GeV and two stau mass eigenstates, $\tilde\tau_1$ and $\tilde\tau_2$, are $\sim 300$ GeV and $\sim 670$ GeV, respectively. The heavy spectrum includes the second chargino, the squarks, the gluino, the lightest stop and the CP even Higgs $H_0$. Note that the charged and CP odd Higgses have masses comparable to $H_0$. 
  
It is also of interest to investigate the effect of the $g-2$ constraint on specific grand unified models.
As a model example, we consider the case of an $\mathsf{SO(10)}$ model where the doublet-triplet 
 mass splitting is obtained in a natural fashion via the missing partner mechanism. For $\mathsf{SU(5)}$ 
  this is accomplished in~\cite{Masiero:1982fe,Grinstein:1982um} while for $\mathsf{SO(10)}$ in~\cite{Babu:2006nf,Babu:2011tw,Aboubrahim:2020dqw,Aboubrahim:2021phn} (for a review of 
  other grand unified models see~\cite{Nath:2006ut}).
The $\mathsf{SO(10)}$ model also allows for a $b-t-\tau$ unification~\cite{Ananthanarayan:1992cd,Chattopadhyay:2001va} along with other desirable properties.
  A neural network analysis again leads us to conclusions similar to the above, i.e., that the radiative electroweak symmetry breaking is gluino-driven. Further, we note that in the analyses of  the cases discussed above, the light sparticle spectrum falls in three
  classes of mass hierarchies which we label as (A), (B) and (C)  
  as  given below.
\\~\\
\noindent
\textbf{(A):}  Here
$\tilde\chi^0_2,\tilde\chi^{\pm}_1$ are essentially degenerate, and $\tilde \tau_1$ is the next lightest 
  supersymmetric particle (NLSP). This leads to the mass hierarchy 
$$m_{\tilde\tau_1}<m_{\tilde\chi^0_2}, 
m_{\tilde\chi^{\pm}_1}<m_{\tilde\ell},$$
where $\tilde\ell$ stands for selectron or smuon.
\\~\\
\textbf{(B):} Here there is a reversal in the hierarchy for the first two inequalities, i.e., between $\tilde \tau_1$
and  $\tilde\chi^0_2$ or $\tilde\chi^{\pm}_1$ which leads to the following possibilities 
\begin{align*}
&m_{\tilde\chi^0_2},  m_{\tilde\chi^{\pm}_1}<m_{\tilde\tau_1}<m_{\tilde\ell},\\
&m_{\tilde\chi^{\pm}_1}<m_{\tilde\chi^0_2}<m_{\tilde\tau_1}<m_{\tilde\ell},\\
&m_{\tilde\chi^{\pm}_1}<m_{\tilde\tau_1}<m_{\tilde\chi^0_2}<m_{\tilde\ell}.
\end{align*}
\\
\textbf{(C):} In this case the selectron and the smuon are lighter than the chargino and the second neutralino 
while the stau is the NLSP. Thus here we have
\begin{equation*}
m_{\tilde\tau_1}<m_{\tilde\ell}< (m_{\tilde\chi^0_2},  m_{\tilde\chi^{\pm}_1}).
\end{equation*}

The Fermilab result also puts constraints on the  allowed region of CP phases arising from the
soft parameters specifically the gaugino masses. Thus the gaugino masses $m_1$ and $m_2$ 
can be complex and one may write $m_1=|m_1| e^{i\xi_1}$ and $m_2= |m_2|e^{i\xi_2}$.
It turns out that the electroweak corrections to the muon $g-2$ are very sensitive to the CP phases~\cite{Ibrahim:1999aj,Ibrahim:2001ym,Aboubrahim:2016xuz}.
As a consequence the measured \amu~puts rather stringent constraints on the allowed range of 
 the CP phases. 
  It is to be noted that the allowed region is further
 constrained by the experimental limits on the EDM of the electron and of the neutron and by the  
 cancellation mechanism~\cite{Ibrahim:1998je}. The CP phases
 can also affect dark matter analyses (see, e.g.,~\cite{Chattopadhyay:1998wb}) and proton stability~\cite{Ibrahim:2000tx} as well as the production cross sections and decays of the sparticles.
 However, an investigation of these is outside the scope of this work
 but these effects are  worth investigation in the future in view of the current experimental result on
 the possible deviation \amu~from the Standard Model prediction.

\section{Cosmologically consistent $\tilde g$REWSB models with split squark-slepton spectra}\label{sec:split}

Of the many parameter points satisfying all the constraints including the recent muon $g-2$ result, we select a few benchmarks to investigate the potential of discovering the low-lying SUSY spectrum at the LHC. In particular we look at the production of light sleptons and sneutrinos and their decay at the HL-LHC and HE-LHC. The benchmarks are given in Table~\ref{tab1} and arranged according to the categories (A), (B) and (C) described earlier. Thus benchmark (I) belongs to case (A), benchmarks (II) and (III)
belong to case (B) and benchmark (IV) 
belongs to case (C). The benchmarks have an $m_0$ of $\mathcal{O}(100)$ GeV while  $m_3$ lies in 
   the several TeV range consistent with a gluino-driven radiative electroweak symmetry breaking. It is noted that benchmark (II)  is taken from Ref.~\cite{Aboubrahim:2021rwz} (labeled there as (d)) 
   while 
   (I), (III), (IV) 
   are from Ref.~\cite{Aboubrahim:2021phn} (labeled there as 
   (f), (i), (h),
    respectively). 
    
\begin{table}[!htp]
\centering
\caption{\label{tab1}
$\tilde g$SUGRA benchmarks for future SUSY searches. All masses are in GeV.}
\begin{tabular}{|ccccccc|}
\hline\hline
Model & $m_0$ & $A_0$ & $m_1$ & $m_2$ & $m_3$ & $\tan\beta$ \\
   \hline\rule{0pt}{3ex}
\!\!(I)  & 688 & 1450 & 852 & 634 & 8438 & 16.8 \\
 (II) & 389 & 122 & 649 & 377 & 4553 & 8.2 \\ 
 (III) & 452 & 648 & 624 & 346 & 4843 & 13.1\\
 (IV)  & 206  & 603 & 842 & 1298 & 7510 & 8.0 \\
\hline\hline 
\end{tabular}
\end{table}

We exhibit in Table~\ref{tab2} the low-lying sparticle mass spectrum for the selected SUGRA benchmarks. The Higgs boson mass satisfies the constraint $125\pm 2$ GeV (allowing for theoretical uncertainties) and the relic density is below the experimental limit which would allow for a multicomponent DM scenario. Also shown is the muon $g-2$ calculated at the two-loop level using \code{GM2Calc}~\cite{Athron:2015rva}. We have also imposed the LHC limits on sparticle masses and constraints on the proton-neutralino spin-independent cross section. Notice that the left handed slepton is lighter than the right handed one except in benchmark (IV) due to $m_2$ being significantly greater that $m_1$ as seen in Table~\ref{tab1}. 

\begin{table}[!htp]
\centering
\caption{\label{tab2}
The SM-like Higgs mass, the light sparticle spectrum and the dark matter relic density $\Omega h^2$ for the benchmarks of Table \ref{tab1}. Also shown is the muon $g-2$. }
\begin{tabular}{|cccccccccc|}
\hline\hline
Model & $h^0$  & $\tilde\ell_{\rm L}$ & $\tilde\ell_{\rm R}$ & $\tilde\nu_{\rm L}$ & $\tilde\tau_1$ & $\tilde\chi^0_1$ & $\tilde\chi^{\pm}_1$ & $\Omega h^2$ & $\Delta a_{\mu}(\times 10^{-9})$ \\
\hline\rule{0pt}{3ex}
\!\!(I)  & 123.0 & 508.1 & 762.0 & 502.3 & 331.9 & 324.2 & 404.3 & 0.004 & 2.11 \\
  (II) & 123.4 & 305.0 & 463.0 & 295 & 251.7 & 237.4 & 237.6 & 0.002 & 2.33 \\
 (III)  & 123.7 & 346.8 & 511.9 & 338.0 & 240.3 & 205.6 & 205.8 & 0.001 & 2.67 \\
 (IV)  & 124.5 & 628.7 & 402.2 &  623.6  & 338.3 & 326.8 & 998.4  & 0.082 & 1.94 \\
\hline\hline
\end{tabular}
\end{table}

For the collider analysis part, we focus on three main SUSY production channels: slepton and sneutrino pair production and slepton associated production with a sneutrino. The latter has a significant cross section since it proceeds via the charged current. The production cross sections are known at the aNNLO+NNLL accuracy and for that we use \code{Resummino-3.0}~\cite{Debove:2011xj,Fuks:2013vua} to calculate the production cross sections at 14 TeV and 27 TeV. The results are shown in Table~\ref{tab3}, where one can see the importance of including the right handed slepton for benchmark
(IV) 
since in this case $m_{\tilde\ell_{\rm R}}<m_{\tilde\ell_{\rm L}}$. 

\begin{table}[!htp]
\centering
\caption{\label{tab3}
The aNNLO+NNLL production cross-sections of slepton pair $(\tilde \ell= \tilde e, \tilde \mu)$ and sneutrino pair as well as slepton associated production with a sneutrino  at $\sqrt{s}=14$ TeV and $\sqrt{s}=27$ TeV for the benchmarks of 
Table \ref{tab1}. The cross section is in fb.} 
\begin{tabular}{|ccc|cc|cc|cc|}
\hline\hline
Model & \multicolumn{2}{c|}{$\sigma(pp\rightarrow \tilde\ell_{\rm L}\,\tilde\ell_{\rm L})$} & \multicolumn{2}{c|}{$\sigma(pp\rightarrow \tilde\ell_{\rm R}\,\tilde\ell_{\rm R})$} & \multicolumn{2}{c|}{$\sigma(pp\rightarrow \tilde\nu_{\rm L}\,\tilde\ell_{\rm L})$}& \multicolumn{2}{c|}{$\sigma(pp\rightarrow \tilde\nu_{\rm L}\,\tilde\nu_{\rm L})$} \\
\hline
&14 TeV & 27 TeV & 14 TeV & 27 TeV & 14 TeV & 27 TeV & 14 TeV & 27 TeV \\
\hline \rule{0pt}{3ex}
\!\!(I) & 1.084 & 4.515 & 0.057 & 0.335 &   &   &  &    \\
(II) & 9.810 & 30.80 & 0.656 & 2.54 & 41.78 & 127.80 & 10.52 & 33.10 \\
(III) & 5.805 & 19.31 & 0.417 & 1.72 & 24.40 & 78.97 & 4.94 & 15.10 \\
(IV) & 0.388 & 1.915 & 1.22 & 4.32  &   &  &  &  \\
  \hline
\end{tabular}
\end{table}

In SUGRA high scale models, the final states resulting from the decay of sleptons and sneutrinos are very rich which is in contrast to simplified models where unit branching ratios are used. The different decay channels for our benchmarks are displayed in Table~\ref{tab4}. Since LHC analyses use simplified models, exclusion limits on sparticle masses do not apply directly to high scale models, but rather limits on $\sigma\times\text{BR}$ become the relevant ones.  

\begin{table}[!htp]
\centering
\caption{\label{tab4}
The branching ratios of relevant decay channels of the left and the right handed slepton and the sneutrino for the benchmarks of Table \ref{tab1}. } 
\begin{tabular}{|ccccccc|}
\hline\hline\rule{0pt}{3ex}
Model & $\tilde\ell_{\rm L}\to \ell\tilde\chi^0_1$ & $\tilde\ell_{\rm L}\to \ell\tilde\chi^0_2$ & $\tilde\ell_{\rm L}\to \nu_{\ell}\tilde\chi^{\pm}_1$ & $\tilde\ell_{\rm R}\to \ell\tilde\chi^0_1 [\tilde\chi^0_2]$ & $\tilde\nu_{\rm L}\to \tilde\chi^+_1\ell^-$ & $\tilde\nu_{\rm L}\to \tilde\chi^0_1\nu_\ell$ \\
\hline \rule{0pt}{3ex}
\!\!(I) & 22\% & 26\% & 52\% & 100\% [-] & 51\% & 24\%  \\
(II) & 31\% & 7\% & 62\% & - [100\%] & 62\% & 30\%  \\
(III) & 31\% & 6\% & 63\% & - [100\%] & 62\% & 31\% \\
(IV) & 100\% & - & - & 100\% [-] & - & 100\% \\
  \hline
\end{tabular}
\end{table}

\section{Detection of sparticles using deep neural network}\label{sec:dnn}

Searches at the LHC target specific final states which are the result of the decay of SUSY particles as shown in Table~\ref{tab4}. Hadronic final states often have a larger branching ratio but this advantage is offset by the presence of a large QCD background, while this problem is significantly less for leptonic final states albeit the small BR. For our analysis we consider final states consisting of two leptons of the same flavor and opposite sign (SFOS) and missing transverse energy (MET) due to the neutralinos (and neutrinos). Jets also play a role in designing the signal regions (SR), where in one SR we require exactly one non-b-tagged jet and in the other we require at least two non-b-tagged jets. Such final states can be easily attributed to slepton pair production but it may be less clear as to why sneutrino pair production and slepton-sneutrino associated production could produce the same final states. As shown in Table~\ref{tab3}, such production channels are relevant for benchmarks (II) and (III). Those benchmarks belong to the mass hierarchy case (B) where the chargino is nearly mass degenerate with the LSP. Because of this, a chargino decay may escape detection at the LHC due to its soft decay products which makes it behave as missing energy. With a 62\% branching ratio for $\tilde\nu_{\rm L}\to\tilde\chi^+_1\ell^-$ (see Table~\ref{tab4}), sneutrino pair production will result in the same final states with a significant $\sigma\times\text{BR}$. The same applies for slepton associated production with a sneutrino which has the largest cross section among the different production channels considered.    

For the final states under consideration, the dominant SM backgrounds are from diboson production, $Z/\gamma+$jets, dilepton production from off-shell vector bosons ($V^*\rightarrow\ell\ell$), $t\bar{t}$ and $t+W/Z$. The subdominant backgrounds are Higgs production via gluon fusion ($ggF$ H) and vector boson fusion (VBF). The signal, which  involves the different production channels presented in Table~\ref{tab3}, and the SM background events are simulated at LO with \code{MadGraph5}~\cite{Alwall:2014hca} and showered with \code{PYTHIA8}~\cite{Sjostrand:2014zea} with the addition of ISR and FSR jets. Cross sections are scaled to their NLO values for the background and to aNNLO+NNLL for the signal. Detector effects are added by the help of \code{DELPHES-3.4.2}~\cite{deFavereau:2013fsa}. 

The standard technique in LHC analyses uses a set of kinematic variables with a high separation power between the signal and the background. A cut-and-count analysis aims at removing as much background as possible while retaining as many of the signal events. Modern techniques involve the use of machine learning such as boosted decision trees or neural networks as an example. A neural network uses a set of kinematic variables to train on the signal and background samples and applies what was `learnt' on a statistically independent set of events. The result is a new variable which we can use as a powerful discriminating variable. In our analysis, we use a deep neural network (DNN) which is part of the `Toolkit for Multivariate Analysis' (TMVA)~\cite{Speckmayer:2010zz} framework within \code{ROOT6}~\cite{Antcheva:2011zz}. The set of kinematic variables used to train the DNN include:
\begin{enumerate}

\item $E^{\rm miss}_{\rm T}$: the total missing transverse energy in an event.

\item $p_{\rm T}(j_1)$: the transverse momentum of the leading non-b-tagged jet.

\item $p_{\rm T}(\ell_1)$ and $p_{\rm T}(\ell_2)$: the transverse momenta of the leading and subleading lepton in an event. 

\item $p_{\rm T}^{\rm ISR}$: the total transverse momentum of the ISR jets in an event. 

\item $M_{\rm T2}$, the stransverse mass~\cite{Lester:1999tx, Barr:2003rg, Lester:2014yga} of the leading and subleading leptons
\begin{equation}
    M_{\rm T2}=\min\left[\max\left(m_{\rm T}(\mathbf{p}_{\rm T}^{\ell_1},\mathbf{q}_{\rm T}),
    m_{\rm T}(\mathbf{p}_{\rm T}^{\ell_2},\,\mathbf{p}_{\rm T}^{\text{miss}}-
    \mathbf{q}_{\rm T})\right)\right],
    \label{mt2}
\end{equation}
where $\mathbf{q}_{\rm T}$ is an arbitrary vector chosen to find the appropriate minimum and the transverse mass $m_T$ is given by 
\begin{equation}
    m_{\rm T}(\mathbf{p}_{\rm T1},\mathbf{p}_{\rm T2})=
    \sqrt{2(p_{\rm T1}\,p_{\rm T2}-\mathbf{p}_{\rm T1}\cdot\mathbf{p}_{\rm T2})}.
\end{equation}  

\item The minimum of the transverse mass $M^{\rm min}_{\rm T}$ defined as $M^{\rm min}_{\rm T}=\text{min}[m_{\rm T}(\textbf{p}_{\rm T}^{\ell_1},\textbf{p}^{\rm miss}_{\rm T}),m_{\rm T}(\textbf{p}_{\rm T}^{\ell_2},\textbf{p}^{\rm miss}_{\rm T})]$. 
The variables $M_{\rm T2}$ and $M^{\rm min}_{\rm T}$ has more discriminating power especially for events with large MET.

\item The dilepton invariant mass, $m_{\ell\ell}$, helps in rejecting the diboson background by requiring $m_{\ell\ell}>100$ GeV, especially near the $Z$ boson pole mass. 

\item The opening angle between the MET system and the dilepton system, $\Delta\phi(\textbf{p}_{\rm T}^{\ell},\textbf{p}^{\rm miss}_{\rm T})$, where $\textbf{p}_{\rm T}^{\ell}=\textbf{p}_{\rm T}^{\ell_1}+\textbf{p}_{\rm T}^{\ell_2}$. 

\item The smallest opening angle between the first three leading jets in an event and the MET system, $\Delta\phi_{\rm min}(\textbf{p}_{\rm T}(j_i),\textbf{p}^{\rm miss}_{\rm T})$, where $i=1,2,3$.

\end{enumerate}

Jets are classified as ISR and non-ISR according to the following recipe: after reconstructing the momentum of the dilepton system, we determine the angle between the dilepton system and each non-b-tagged jet in the event. We select up to two jets that are closest to the dilepton system and tag them as possible jets arising from the decay of the SUSY system while the rest are classified as ISR jets. 
\begin{table}[!htp]
\caption{The preselection criteria and the analysis cuts on a set of kinematic variables at 14 TeV (27 TeV) grouped by the benchmarks of Table~\ref{tab1} in two signal regions SR-$2\ell1$j and SR-$2\ell2$j. Entries with a dash (-) imply  that no requirement on the variable is considered.  Cuts are optimized for each center-of-mass energy. }
\resizebox{\textwidth}{!}{\begin{tabular}{|c|ccc|ccc|}
\hline\hline
\multirow{2}{*}{Observable}  
& (II), (III) & (I) & (IV)  & (II), (III) & (I) & (IV)  \\
\cline{2-7}\rule{0pt}{3ex}
 & \multicolumn{3}{c|}{Preselection criteria (SR-$2\ell2$j)} & \multicolumn{3}{c|}{Preselection criteria (SR-$2\ell1$j)} \\
 \hline \rule{0pt}{3ex}
 $N_{\ell}$ (SFOS) & \multicolumn{3}{c|}{$2$} & \multicolumn{3}{c|}{$2$} \\
$N_{\rm jets}^{\rm non-b-tagged}$ & \multicolumn{3}{c|}{$\geq 2$} & \multicolumn{3}{c|}{$1$}  \\
$p_T(j_1)$ [GeV] & \multicolumn{3}{c|}{$>20$} & \multicolumn{3}{c|}{$>20$} \\
$p_T(\ell_1)$ (electron, muon) [GeV] & \multicolumn{3}{c|}{$>15$, $>10$} & \multicolumn{3}{c|}{$>15$, $>10$} \\
$E^{\rm miss}_T$ [GeV] & \multicolumn{3}{c|}{$>100$} & \multicolumn{3}{c|}{$>100$}\\
\cline{2-7}\rule{0pt}{3ex}
 & \multicolumn{3}{c|}{Analysis cuts} & \multicolumn{3}{c|}{Analysis cuts} \\
\cline{2-7}\rule{0pt}{3ex}
\!\! $m_{\ell\ell}~\text{[GeV]} >$ & 136 (110) & 150  & 150 (110) & 110 & 120  & 150   \\
 $E^{\rm miss}_T/\textbf{p}^{\ell}_{\rm T}>$ & 1.9 (2.8)  &  -  & -  & 1.0  &  -  & -   \\
$\Delta\phi_{\rm min}(\textbf{p}_{\rm T}(j_i),\textbf{p}^{\rm miss}_{\rm T})~\text{[rad]} >$ & -  & 0.85 (1.5)  &  -  & -  & 0.85 (1.5)  &  - \\
$p_T(\ell_2)~\text{[GeV]} >$ & -  &  -  & 190 (370) & -  &  -  & 190 (300)   \\
$M_{T2}~\text{[GeV]} >$ & - (140)  &  - (120)  & 200 (300) & 130 (230)   & 100  & 200 (300)  \\
DNN response $>$   & 0.95  & 0.95 & 0.95 & 0.95  & 0.95 & 0.95 \\
\hline\hline
\end{tabular}}
\label{tab5}
\end{table}
We summarize in Table~\ref{tab5} the preselection criteria as well as the analysis cuts used on the signal and background events.  The cuts are optimized depending on whether they belong to the mass categories (A), (B) or (C) discussed earlier and on the center-of-mass energy. 

\begin{figure}[!htp]
\centering
\includegraphics[width=0.49\textwidth]{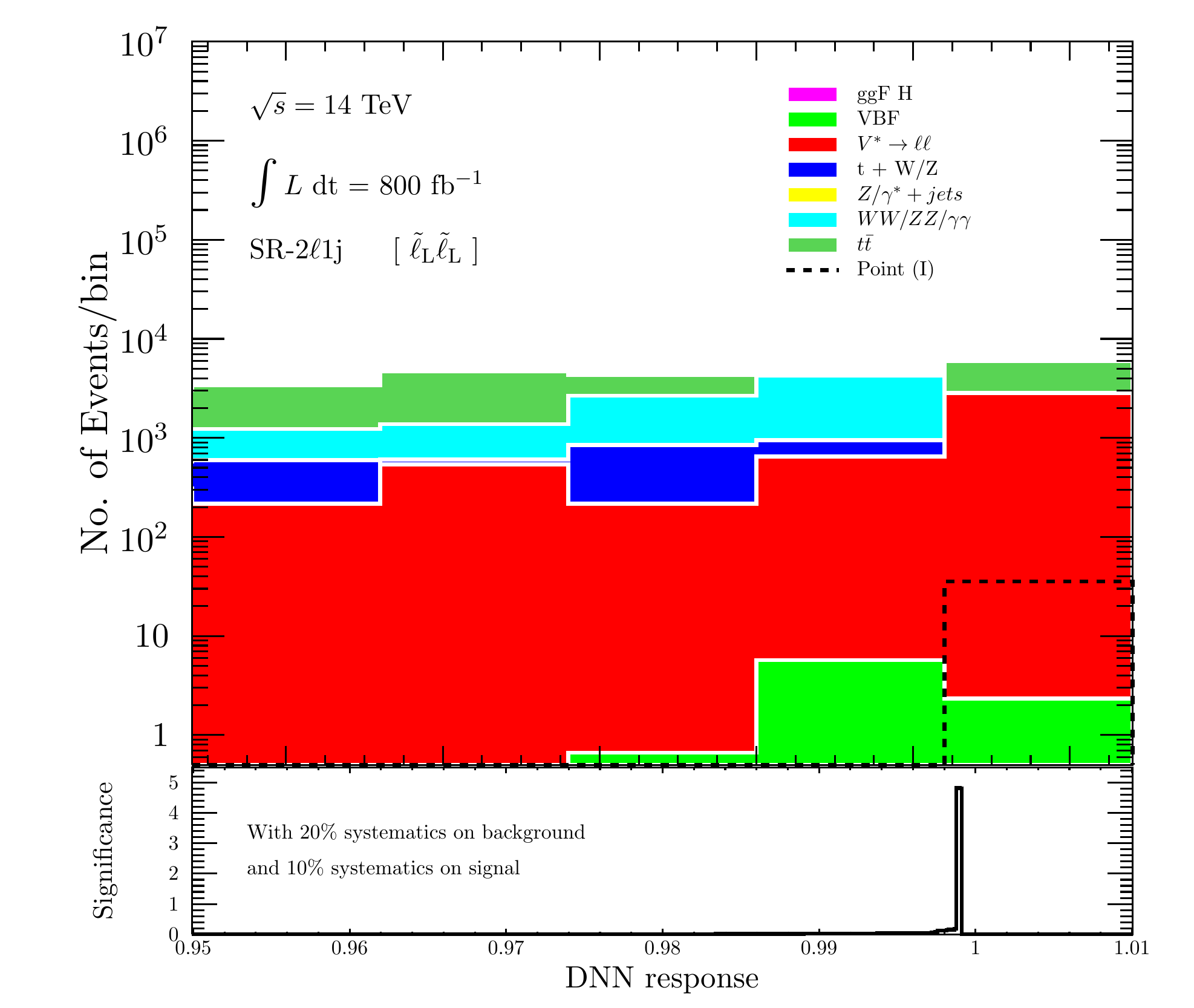}
\includegraphics[width=0.49\textwidth]{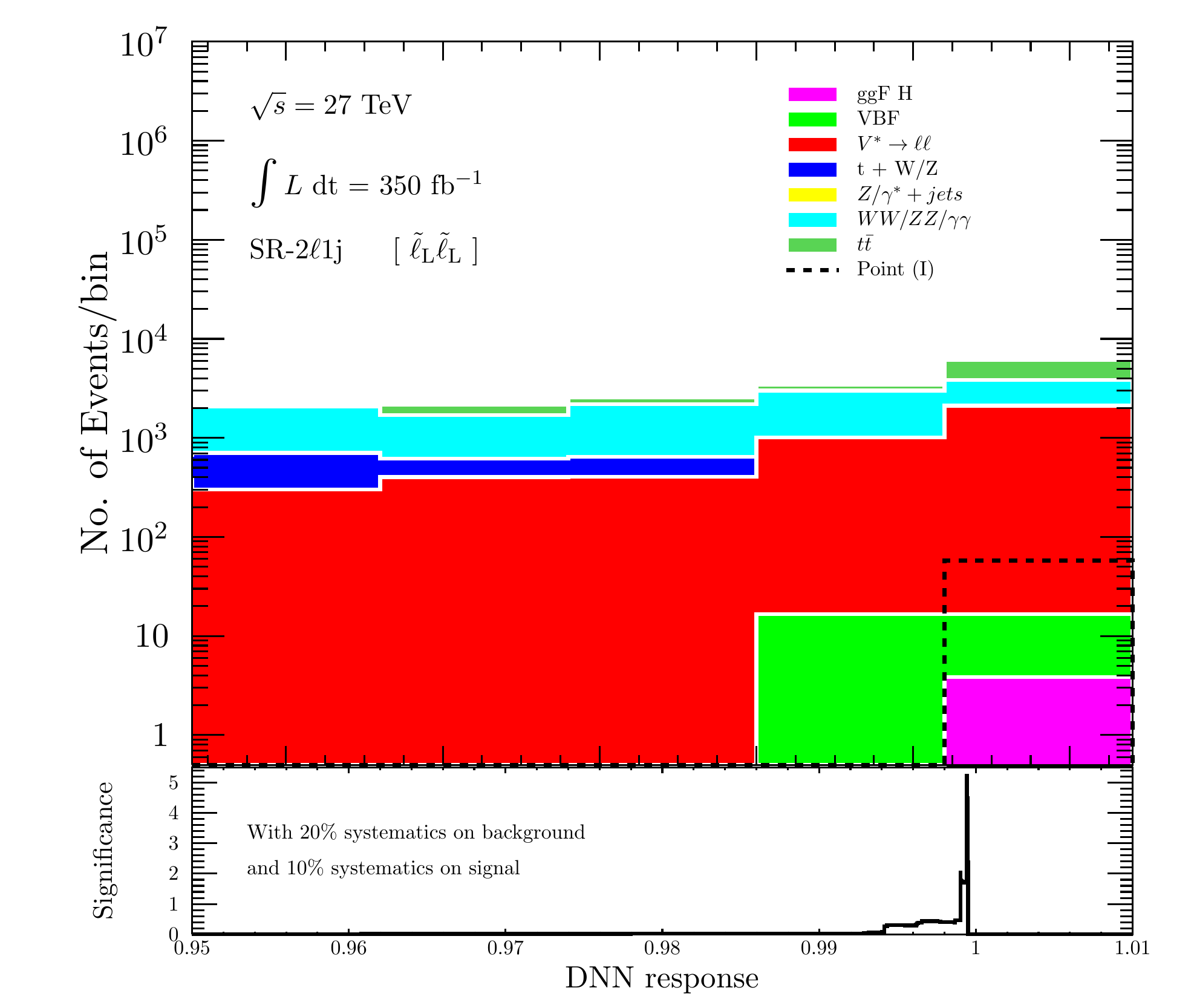} 
\caption{Distributions in the `DNN response' variable for
benchmark (I)
at 14 TeV (left) and 27 TeV (right). A distribution in the signal significance as a function of the cut on the `DNN response' is shown in the bottom pads of each panel. } 
\label{dnn_dist}
\end{figure}

As mentioned before, the end result of the training and testing stages by the DNN is a new discriminating variable called the `DNN response' which tends to take values closer to one for the signal. Signal (S) and background (B) distributions in the `DNN response' variable are shown in Fig.~\ref{dnn_dist} for 
for benchmark (I)  at 14 TeV and 27 TeV. In the bottom pad of each panel, we show the distribution in the signal significance $Z$ as a function of the cut on `DNN response' and taking into consideration the systematics, i.e.,
\begin{align}
Z=\frac{S}{\sqrt{S+B+(\delta_S S)^2+(\delta_B B)^2}},
\label{significance}
\end{align}
where $\delta_S$ and $\delta_B$ are the systematic uncertainties in the signal and background estimates. The recommendations on systematic uncertainties (known as `YR18' uncertainties) published in the CERN's yellow reports~\cite{CidVidal:2018eel,Cepeda:2019klc} suggest an overall 20\% uncertainty in the background and 10\% in the SUSY signal. Notice that a $5\sigma$ value is reached at the integrated luminosities shown which can be attained at HL-LHC and HE-LHC.

Of the two signal regions considered, we find that the most optimal one is the single jet signal region, which has also been shown to be true in previous LHC searches~\cite{Aad:2019vnb}. This is evident from Table~\ref{tab6} where the integrated luminosity for discovery of the benchmarks in SR-$2\ell2$j is much larger than in SR-$2\ell1$j after combining all production channels. The presented integrated luminosities include the effect of systematics.   

\begin{table}[!htp]
\centering
\caption{\label{tab6} The estimated integrated luminosities, in fb$^{-1}$,  for discovery of benchmarks of Table~\ref{tab1} at 14 TeV and 27 TeV after combining all production channels and including systematics in the signal and background. } 
\resizebox{0.75\textwidth}{!}{\begin{tabular}{|ccc|cc|}
\hline\hline
Model & \multicolumn{2}{c|}{SR-$2\ell1$j}& \multicolumn{2}{c|}{SR-$2\ell2$j}  \\
\hline \rule{0pt}{3ex}
& $\mathcal{L}$ at 14 TeV & $\mathcal{L}$ at 27 TeV & $\mathcal{L}$ at 14 TeV & $\mathcal{L}$ at 27 TeV\\
\hline \rule{0pt}{3ex}
\!\!(I) & 880 & 310 & 1262 & 694  \\
(II) & 200 & 50 & 1860 & 715  \\
(III) & 148 & 75 & 1887 & 1320  \\
(IV) & 1040 & 232 & 1738 & 1194  \\
 \hline\hline
\end{tabular}}
\end{table}

\begin{figure}[!htp]
\centering
\includegraphics[width=0.75\textwidth]{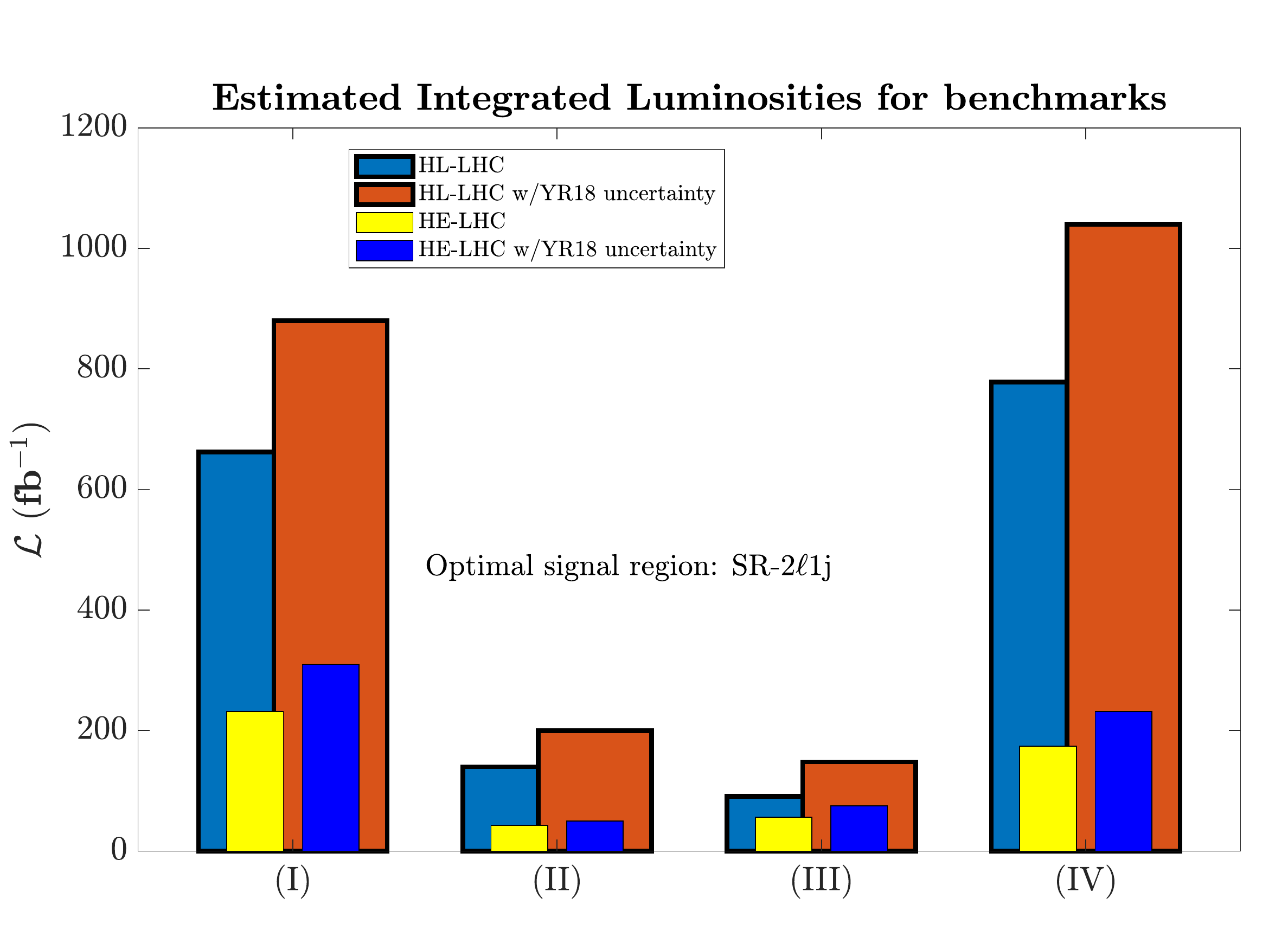}
\caption{The integrated luminosities, $\mathcal{L}$, needed for discovery of  SUSY at HL-LHC and HE-LHC assuming that \amu~arises from SUSY loops. Values of $\mathcal{L}$ are shown before and after including the `YR18' uncertainties on the signal and background.} 
\label{figlumi}
\end{figure}

As a comparison between HL-LHC and HE-LHC and to see the effect of systematic uncertainties, we plot the estimated integrated luminosities for discovery of benchmarks  (I)$-$(IV) 
at 14 TeV and 27 TeV in Fig.~\ref{figlumi}. Benchmarks 
(II) and (III) 
require 200 fb$^{-1}$ and 148 fb$^{-1}$ at 14 TeV which should be attained in the coming run of LHC. The rest require more than  400 fb$^{-1}$ but are all within the reach of HL-LHC. The same benchmarks require much smaller integrated luminosities for discovery at HE-LHC. 

One final remark regarding the mass spectrum shown in Table~\ref{tab2}. Benchmarks (I) and (IV) 
are characterized by a light weakino spectrum with a considerable mass gap between the charginos and neutralinos. Thus it is imperative that one looks at the weakino production channel. It turns out that in those benchmarks, the stau is the NLSP
and
 not the chargino (or the second neutralino). The stau will then decay to a tau and the LSP, where the tau can decay leptonically but with a small branching ratio. Overall, the targeted final states of SFOS leptons seem to have a much smaller $\sigma\times$BR compared to slepton and sneutrino production and so such a production mechanism is not significant in our analysis.  
We note that recently several works 
 within SUSY have come out regarding an explanation of the Fermilab muon
anomaly~\cite{Iwamoto:2021aaf,Gu:2021mjd,VanBeekveld:2021tgn,Yin:2021mls,Wang:2021bcx,Cao:2021tuh,Chakraborti:2021dli,Han:2021ify,Baum:2021qzx,Ahmed:2021htr,Baer:2021aax,Endo:2021zal,Ibe:2021cvf,Chakraborti:2021bmv,Athron:2021iuf,Borah:2021jzu,Altmannshofer:2021hfu,Zheng:2021wnu,Jeong:2021qey,Ellis:2021zmg,Frank:2021nkq,Abdughani:2021pdc,Abdughani:2019wai,Li:2021bbf,Wang:2021uqz}.

\section{Conclusion \label{sec:conc}}

The recent muon $g-2$ result from the Fermilab Collaboration confirms the earlier result from the Brookhaven experiment regarding a deviation from the Standard Model prediction. Specifically, the combined Fermilab and Brookhaven results give a $4.1\sigma$ deviation from the Standard Model prediction compared to $3.7\sigma$ deviation for Brookhaven which supports the Brookhaven result and  strengthens it. The Fermilab result can be explained in supersymmetry
if the sleptons and some of the weakino masses are low-lying in
the range of few hundred GeV. On the other hand, the observation
of the Higgs boson mass at 125 GeV requires that the mass scale for the
squarks be much larger and lie in the few TeV region which indicates
a wide mass gap between the squark masses and the slepton 
masses. Such a mass gap can be generated in a natural way 
in gluino-driven radiative breaking of the electroweak symmetry
where the color interactions of the gluino drive the masses
of the quarks to high values while the masses of the sleptons
can be low if the universal scalar mass is chosen to be small.
The same 
result is arrived at using an artificial neural network where the
SUGRA parameter space is scanned to determine the region
where a large squark/slepton mass hierarchy can arise and the
analysis again leads to gluino-driven radiative breaking.
       
        The existence of light sparticle spectrum consisting of sleptons
        and some of the electroweak gauginos, opens up the possility of 
        their discovery at the LHC as well as a test of $\tilde g$SUGRA.      For this purpose, we train a deep neural network on the signal and background events to generate a discriminating variable capable of increasing the signal to background ratio. We exhibit the importance of this technique 
       by analyzing the integrated luminosities needed for the discovery of the  benchmarks, and show that some of the benchmarks can be discovered 
          at  HL-LHC with integrated luminosities as low as 150$-$300 fb$^{-1}$ while much smaller integrated luminosities are required at HE-LHC. We propose that it would be fruitful if the future SUSY analyses at the LHC are carried out with the inclusion of the Fermilab $g-2$ constraint  and further such analyses need to go beyond generic simplified models to well motivated high scale models.    Thus dedicated studies for the production and signal analysis of sleptons and weakinos in $\tilde g$SUGRA in RUN-3 at the LHC 
               is indicated.

\section*{Acknowledgments}

The research of AA and MK was supported by the BMBF under contract 05H18PMCC1. The research of PN  was supported in part by the NSF Grant PHY-1913328.

\section*{References}

\end{document}